\documentclass[11pt,aps,prl,tightenlines]{revtex4}
\usepackage{epsfig,amsmath}

\newcommand{\be}{\begin{equation}}
\newcommand{\ee}{\end{equation}}
\newcommand{\bea}{\begin{eqnarray}}
\newcommand{\eea}{\end{eqnarray}}
\newcommand{\fb}{{\cal F}_B}
\newcommand{\fbpa}{{\cal F}_{B \parallel}}
\newcommand{\fbpe}{{\cal F}_{B \perp}}

\begin{document}

\hspace{13.2cm}{$\,$TUM/T39-02-14}

\hspace{14.cm}{ECT*-02-17}
\vspace{5mm}

\title{\Large Bound state kinetics in high-energy nuclear collisions}
\author{Alberto Polleri}
\affiliation{Physik Department, Technische Universit\"{a}t M\"{u}nchen,
D-85747 Garching, GERMANY\\
ECT*, Villa Tambosi, I-38050 Villazzano (Trento), ITALY}
\begin{abstract}
A Lorentz covariant kinetic equation for bound states and their constituents
is presented and solved exactly in closed form. It describes in a unified way
dynamical formation and dissociation of states such as quarkonia and 
(anti)-deuterons in the excited medium formed with a high-energy heavy-ion
collision.
\end{abstract}
\maketitle 

The possibility to describe hadronic bound states in strongly
interacting matter provides an important tool for understanding the
dynamics of the medium produced in high-energy heavy-ion collisions.
The main reason lies in the fact that bound states can be formed and
destroyed during specific stages of evolution of the medium, with a
delicate interplay among various elements: the binding energy and structure
of the object under consideration, the medium nature and characteristics
which change as function of time and the abundances and phase-space
correlations of the bound state constituents.

There are essentially two important kinds of bound states which
deserve attention: heavy quarkonia ($J/\psi$, $\Upsilon$ and excited
states) and (anti-)deuterons ($d$, $\bar d$). Their
properties are quite different from one another and for this reason
they probe different stages of a nuclear collision. The former are
tightly bound, small objects, in general characterized by a hard
scale and subject to confinement. They probe the early stages of a 
collision, in particular a possible 
quark-gluon plasma (QGP). The latter are very loosely bound, large
objects which can be formed and survive dissociation only at the very
latest moments of a collision. They probe the freeze-out stage.

Despite the aforementioned differences, the essential physics which 
regulates the time evolution in an interacting medium is very
similar. For both bound states it is possible that they are formed in
the evolving medium and are subsequently dissociated. How this happens
is regulated by transition probabilities and by the medium
constituents' phase-space density. It is therefore desirable to have a 
unified description of these phenomena.

In the following we discuss a general kinetic equation, appropriate
for both cases, and we present in closed form the exact solution,
examining in detail specific features of the two cases under
study. The solution is constructed for two different ways of
specifying the initial condition. Finally, we discuss a simple
analytical example, leaving the more realistic cases for future
numerical calculations.

\begin{figure}[t]
\begin{center}
\epsfig{file=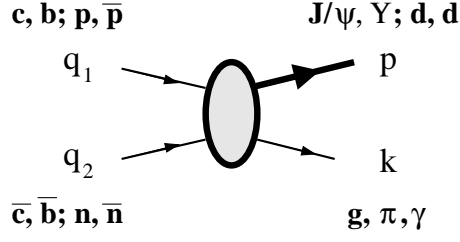,width=6cm}
\caption{Diagram illustrating the amplitude for the formation of a bound
state. The dissociation amplitude is obtained with detailed balance. Possible
reactions are $c \bar c \leftrightarrow J/\psi g (\pi)$,
$b \bar b \leftrightarrow \Upsilon g (\pi)$, 
$pn \leftrightarrow d \pi(\gamma)$,
$\bar p \bar n \leftrightarrow \bar d \pi(\gamma)$}
\label{figure:diagrams}
\end{center}
\end{figure}

We begin by specifying the processes $B(p) + n(k) \leftrightarrow
c_1(q_1) + \bar c_2(q_2)$ of dissociation and formation of a bound
state $B$ into and from a pair of constituents $c_1 c_2$ by means of a
fourth degree of freedom labelled $n$. The corresponding momenta are
indicated in parentheses. As mentioned earlier, the bound state $B$
can be a quarkonium or a(n) (anti-)deuteron, while the respective
constituents are $q \bar q$ pairs of heavy quarks or proton-neutron
$pn$ pairs. In particular, the dominant reactions to be considered are
$c \bar c \leftrightarrow J/\psi g (\pi)$,
$b \bar b \leftrightarrow \Upsilon g (\pi)$, 
$pn \leftrightarrow d \pi(\gamma)$,
$\bar p \bar n \leftrightarrow \bar d \pi(\gamma)$.
The amplitude for the process is illustrated in 
Fig.~\ref{figure:diagrams}. 

Borrowing concepts from the treatment of the non-linear 
Boltzmann equation \cite{GLW80} we postulate that the kinetic equation 
describing the time evolution of
the phase-space density $\fb$ of the bound state is then
\be
p^\mu \partial_\mu\, \fb(p,x) = C_F(p,x) - C_D(p,x)\,\fb(p,x)\,,
\label{equation}
\ee
consisting of a drift term (the differential operator $p^\mu \partial_\mu$)
and a collision term, the latter made of a formation part
\bea
C_F(p,x) \!&=&\!\! \sum_n
\int d\Phi(p,k,q_1,q_2)
\ W_{c_1 c_2 \rightarrow B n}(s,t)\\
&&\hspace{2cm} \times\ f_{c_1}(q_1,x)f_{c_2}(q_2,x) \nonumber
\eea
and a dissociation part
\be
C_D(p,x) = \sum_n
\int d\Phi(p,k,q_1,q_2)
\ W_{B n \rightarrow c_1 c_2}\ f_n(k,x)\,.
\ee
In the above $f_{c_1}, f_{c_2}$ and $f_n$ are the phase-space distributions
of the degrees of freedom participating in the collisions, depending 
implicitly on collision energy and centrality, while the phase-space 
integration measure is
\bea
d\Phi(p,k,q_1,q_2) \!&=&\! \frac{1}{2}\,
\frac{d^3\vec k}{(2\pi)^3 2E_k}\frac{d^3 \vec q_1}{(2\pi)^3 2E_1}
\frac{d^3 \vec q_2}{(2\pi)^3 2E_2}\\
& & \times \ (2\pi)^4 \delta^4(p\!+\!k\!-\!q_1\!-\!q_2)\,. \nonumber
\eea
As evident from the integration measure, all particles are assumed to be on 
their mass shell. The transition probability satisfies detailed balance as
\be
W_{c_1 c_2 \rightarrow B n}(s,t) = W_{B n \rightarrow c_1 c_2}(s,t)
\ee
and a sum over all particle species $n$ constituting the medium and 
contributing to dissociation and formation is indicated. Clearly, which degrees
of freedom are active depends on the evolution time. The arguments 
of the transition probability are the Mandelstam variables $s = (p+k)^2 = 
(q_1+q_2)^2$ and $t = (q_1-p)^2 = (k-q_2)^2$. In the equations above
we have neglected Bose-Einstein or Fermi-Dirac statistics, {\it i.e.} Bose
enhancement and Pauli blocking.

All the dynamics is well defined once one specifies the distributions 
of the medium constituents and those of the bound state constituents,
together with the transition probabilities $W_{c_1 c_2 \leftrightarrow B n}$.
The kinetic equation can then be solved in a closed form, which is
convenient to present in two different ways, depending on how
the initial condition is given, {\it i.e.} either at some initial
global time $t_0$ or at some initial proper time $\tau_0$.
The form of eq.~(\ref{equation}) is very general and in the following we
make no reference to the detailed structure of the collision term. In other
words the functions $C_F$ and $C_D$ are assumed to be known.

When the initial condition is specified at some initial global
time $t_0$ to be $\fb^0(\vec p,\vec r)$, then the kinetic equation is
conveniently re-written as
\bea
\left(\frac{\partial}{\partial t} + \vec v \cdot 
\frac{\partial}{\partial \vec r}\right) 
\, \fb(\vec p,\vec r,t) \!&=&\! \lambda_F(\vec p,\vec r,t) \label{kintime} \\
& & - \, \lambda_D(\vec p,\vec r,t)\ \fb(\vec p,\vec r,t)\,, \nonumber
\eea
where $\vec v = \vec p/E_p$ and $\lambda_{F,D} = C_{F,D}/E_p$.
The solution to this last equation can be obtained with some simple steps.
Neglecting the collision term, it is trivial to see that the solution is
the scaled function
\be
\fb^*(\vec p,\vec r,t) = \fb^0\left(\vec p,\vec \rho(t-t_0) 
\rule{0pt}{10pt} \right)\,,
\ee
where $\rho(t) = \vec r-\vec v\, t$. This is the case of free streaming.
The case with dissociation but without formation was studied in \cite{BO89}
to address the problem of charmonium suppression. There the solution was 
given as
\bea
\fb^{\rm diss}(\vec p,\vec r,t) \!&=&\! 
\fb^*(\vec p,\vec r,t) \label{sol1} \\ 
& & \times\ \exp\left[- \int_{t_0}^t\!dt' 
\lambda_D\!\left(\,\vec p,\vec \rho(t-t'),t' 
\rule{0pt}{10pt} \right) \right] \,. \nonumber
\eea
Otherwise, with the only formation term and without dissociation, the 
solution is obtained by direct integration as
\be
\fb^{\rm form}(\vec p,\vec r,t) = \fb^*(\vec p,\vec r,t)
+\!\! \int_{t_0}^t\!\!\!dt'\, \lambda_F\!\left(\,\vec p,\vec \rho(t\!-\!t'),t'
 \rule{0pt}{10pt} \right)
\label{sol2}
\ee
It is interesting to notice the structure of the scaled argument
$\vec \rho(t-t') = \vec r-\vec v\, (t-t')$, ensuring that any function
of it is annihilated by the drift operator. This observation, although 
trivial, is crucial for the complete solution. We now combine the previous
partial solutions and seek for the general case a solution of the form
\bea
\fb(\vec p,\vec r,t) \!\!&=&\!\!\! \left[
\fb^*(\vec p,\vec r,t)
+ \!\!\int_{t_0}^t\!\!dt'\, \alpha\!\left(\,\vec p,\vec \rho(t\!-\!t'),t'
 \rule{0pt}{10pt} \right)\right] \\
& & \ \ \times\ \exp\left[- \int_{t_0}^t\!dt'\,
\lambda_D\!\left(\,\vec p,\vec \rho(t-t'),t' \rule{0pt}{10pt} \right)
\right]\,,\nonumber
\label{trial}
\eea
where $\alpha$ is a function to be determined.
With this choice, substituting in eq.~(\ref{kintime}) one finds right 
away that the needed function is
\be
\alpha(\vec p,\vec r,t) \!=\! \lambda_F(\vec p,\vec r,t)
\,\exp\!\left[\int_{t_0}^t\!\!dt'
\lambda_D\!\left(\,\vec p,\vec \rho(t\!-\!t'),t'\! \rule{0pt}{10pt} \right)
\!\right] 
\ee
Then, the general solution of the kinetic equation as function of global
time can be obtained by substituting this last result into
the trial form of eq.~(\ref{trial}). The final result is then 
\begin{widetext}
\bea
\fb(\vec p,\vec r,t) \!&=&\! \left\{ \!\rule{0pt}{18pt} \right.
\fb^*(\vec p,\vec r,t)\ \exp\left[- \int_{t_0}^t\!dt'\,
\lambda_D\!\left(\,\vec p,\vec \rho(t-t'),t' \rule{0pt}{10pt} \right)
\right]\label{solglobal}\\ & &\left. \ \ 
+\, \int_{t_0}^t\!dt'\,
\lambda_F\!\left(\,\vec p,\vec \rho(t-t'),t' \rule{0pt}{10pt} \right)
\ \exp\left[- \int_{t'}^{t}\!dt''\,\lambda_D\!
\left(\,\vec p,\vec \rho(t-t''),t'' \rule{0pt}{10pt} \right)\right] 
\right\}\,.\nonumber 
\eea
\end{widetext}
By immediate inspection one finds that the initial condition is
satisfied at $t = t_0$ and that the two previous partial solutions
$\fb^{\rm diss}$ in eq.~(\ref{sol1}) and $\fb^{\rm form}$ in eq.~(\ref{sol2})
are obtained in the respective limiting cases $\lambda_F = 0$ and
$\lambda_D = 0$. We will comment on the obtained result later on, after the
second derivation of the solution

Before we continue note that integrating over the position 
variables gives the invariant spectrum
\be
E_p \frac{dN_B}{d^3\vec p} = \int\!\frac{d^3\vec r}{(2 \pi)^3}
\ \fb(\vec p,\vec r,t_f)
\ee
of the bound state at a given final global time $t_f$.

If the initial condition is specified at some initial proper time $\tau_0$ 
it is convenient to use momentum rapidity 
$y = 1/2\, \log\left(\rule{0pt}{10pt}(E+p_z)/(E-p_z)\right)$
and space-time rapidity
$\eta = 1/2\, \log\left(\rule{0pt}{10pt}(t+z)/(t-z)\right)$
as longitudinal variables. The initial condition is then
$\fb^0(y,\vec p_\perp,\eta,\vec r_\perp)$.
It is then useful to  re-write the drift operator as
$p^\mu \partial_\mu = {\cal D} = {\cal D}_\parallel + {\cal D}_\perp$
where
\bea
{\cal D}_\parallel\!\! &=&\!\! m_\perp \left[ \cosh(y-\eta)\,
\frac{\partial}{\partial \tau} +
\frac{1}{\tau}\, \sinh(y-\eta)\, \frac{\partial}{\partial \eta} \right] \\ 
{\cal D}_\perp \!\!\!    &=&\!\! \vec p_\perp \cdot 
\frac{\partial}{\partial \vec r_\perp}
\eea
As in the previous case, let us first consider the situation of free
streaming, without a collision
term. Neglecting the transverse coordinate dependence for a moment, the 
solution of the free streaming equation with the only operator
${\cal D}_\parallel$ is any function of the variable
\be
\xi = \frac{\tau}{\tau_0}\sinh(y-\eta)\,,
\ee
It is therefore convenient to define the quantity
\be
Y(\xi) = y - \log\left(\xi + \sqrt{1 + \xi^2} \right)\,,
\ee
which reduces to $\eta$ when $\tau = \tau_0$. In this way
one obtains the longitudinal free streaming solution
\be
\fbpa^*(y,\eta,\tau) = \fbpa^0\left(y,Y(\xi) \rule{0pt}{10pt}\right)\,.
\ee
Concerning the transverse part we look for a scaling solution of the type
\be
\fbpe^*(\vec r_\perp,\tau) = \fbpe^0\left(\vec r_\perp - v_\perp \bar\tau 
\rule{0pt}{10pt} \right)\,,
\ee
with $v_\perp = p_\perp / m_\perp$ and $\bar \tau$ a function to be determined.
The kinetic equation becomes
\bea
{\cal D}\fbpe^*  &=&  \frac{\partial}{\partial \vec r_\perp}\,\fbpe^* \cdot (- 
v_\perp) \ {\cal D}_\parallel \,\bar\tau + {\cal D}_\perp \fbpe^* \\
& = & {\cal D}_\perp \fbpe^* \cdot \left(1 - \frac{{\cal D}_\parallel 
\,\bar\tau}{m_\perp}
\right) = 0\,, \nonumber
\eea
which is equivalent, in the non-trivial case ${\cal D}_\perp \fbpe^* \neq 0$,
to the equation ${\cal D}_\parallel\,\bar\tau = m_\perp$.
It is simple to see that the solution is
\be
\bar\tau(\xi,\tau) = \int_{\tau_0}^\tau\!d\tau'\ \frac{1}{\sqrt{1 + 
(\xi \tau_0/\tau')^2}} 
\label{taubar}
\ee
with the correct boundary condition $\bar\tau = 0$ when $\tau = \tau_0$.
The full free streaming solution is obtained by putting together the 
separate results for the longitudinal and transverse parts, yielding
\be
\fb^*(y,\vec p_\perp,\eta,\vec r_\perp,\tau) = 
\fb^0\left(y,\vec p_\perp,Y(\xi),\vec \rho_\perp(\xi,\tau)
\rule{0pt}{10pt} \right)
\ee
being $\vec \rho_\perp(\xi,\tau) = \vec r_\perp - v_\perp \bar\tau(\xi,\tau)$.
We now consider the more complicated case with dissociation, still neglecting
formation. Analogously to the previous case, we look for a solution of the 
form
\bea
\fb^{\rm diss}(y,\vec p_\perp,\eta,\vec r_\perp,\tau) \!&=&\!
\fb^*(y,\vec p_\perp,\eta,\vec r_\perp,\tau) \\
& & \times\ \exp\left[ - E(y,\vec p_\perp,\eta,\vec r_\perp,\tau)
\rule{0pt}{10pt} \right]\,, \nonumber
\eea
where the exponent $E$ is determined by the equation
\be
{\cal D} E(y,\vec p_\perp,\eta,\vec r_\perp,\tau) = 
C_D(y,\vec p_\perp,\eta,\vec r_\perp,\tau)\,.
\ee
The latter can be solved by direct integration, taking care of the fact that
the integration measure should be the same as that of eq.~(\ref{taubar}),
ensuring the correct scaling with the variable $\xi$. The solution is then
\be
E(y,\vec p_\perp,\eta,\vec r_\perp,\tau) =\!\!
\mbox{\Large $\displaystyle \int$}_{\!\!\!\!\!\tau_0}^\tau\!\!\!\!d\tau' 
\frac{C_D\!\!\left(y,\vec p_\perp,Y(\xi \tau_0/\tau'),
\vec \rho_\perp(\xi,\tau') \rule{0pt}{10pt}\! \right)}
{m_\perp\,\sqrt{1 + (\xi \tau_0/\tau')^2}}
\ee
With this last result, especially in its structure concerning the integration
measure, it is now possible to obtain the full solution. Looking at the
earlier solution given with eq.~(\ref{solglobal}) in the previous case
of initialization at fixed global time, one can repeat the 
arguments leading to it and obtain, after defining the rates $\Lambda_{F,D} = 
C_{F,D}/m_\perp$ for dimensional clarity ($[\Lambda_{F,D}] =$ fm$^{-1}$),
the general solution of the kinetic equation as
\begin{widetext}
\bea
\fb(y,\vec p_\perp,\eta,\vec r_\perp,\tau) \!\!&=&\!\! 
\left\{ \rule{0pt}{18pt} \fb^*(y,\vec p_\perp,\eta,\vec r_\perp,\tau) 
\ \exp\!\left[- 
\mbox{\Large $\displaystyle \int$}_{\!\!\!\!\!\tau_0}^\tau\!\!d\tau' 
\ \frac{\Lambda_D\left(y,\vec p_\perp,Y(\xi \tau_0/\tau'),
\vec \rho_\perp(\xi,\tau') \rule{0pt}{10pt} \right)}
{\sqrt{1 + (\xi \tau_0/\tau')^2}}
\right] 
\right.\label{sollocal} \\ 
& & \hspace{-3.5cm} 
\ \, + \!\! \left. 
\mbox{\Large $\displaystyle \int$}_{\!\!\!\!\!\tau_0}^\tau\!\!\!d\tau' 
\ \frac{\Lambda_F\left(y,\vec p_\perp,Y(\xi \tau_0/\tau'),
\vec \rho_\perp(\xi,\tau') \rule{0pt}{10pt} \right)}
{\sqrt{1 + (\xi \tau_0/\tau')^2}}
\ \exp\!\left[
- \mbox{\Large $\displaystyle \int$}_{\!\!\!\!\!\tau'}^{\tau}\!\!\!\!d\tau'' 
\ \frac{\Lambda_D\left(y,\vec p_\perp,Y(\xi \tau_0/\tau''),
\vec \rho_\perp(\xi,\tau'') \rule{0pt}{10pt} \right)}
{\sqrt{1 + (\xi \tau_0/\tau'')^2}}
\right] \!\right\}.
\nonumber
\eea
\end{widetext}
In this case the initial condition is satisfied at $\tau = \tau_0$ and the 
partial solutions $\fb^{\rm diss}$ and $\fb^{\rm form}$ are again recovered
in the respective limiting cases $\Lambda_F = 0$ and $\Lambda_D = 0$.

The structure of the solution is rich of details on the dynamics of
dissociation and formation. The first part of the r.h.s. of
eq.~(\ref{sollocal}), but see also eq.~(\ref{solglobal}), gives the final
phase-space density of those bound states which were initially
present at the start of the evolution at $\tau = \tau_0$. This term is
especially important for the study of quarkonia. These are initially 
produced by hard collisions when the two nuclei first interact and are
subsequently dissociated by the formed medium. On the other hand, at the
high collision energies considered,
(anti-)deuterons are absent at any early stage of a heavy-ion
collision, therefore this term can be neglected. The second piece of the
solution describes formation of bound states from $\tau_0$ up to
$\tau'$ and their subsequent dissociation from $\tau'$ until
$\tau$. For quarkonia it describes formation in a QGP, process that
might be significant when the phase-space occupation of heavy quarks
is large enough. This might happen at high collision energies, as recently
suggested \cite{BMS00,TSR01}. For (anti-)deuterons this second term is
the relevant one, describing coalescence of $pn$ pairs. This process
is relevant only at low enough medium densities (small $\Lambda_D$),
{\it i.e.} at the freeze-out stage. A more sophisticated treatment of
this problem for deuterons at low and intermediate collision energies
was developed in \cite{DB91}. Here, since we consider the high-energy
collision regime, we can safely neglect many-body correlations. Hence
the problem is significantly simpler.

An interesting application of eq.~(\ref{sollocal}) for the study of
quarkonia is the comprehensive analysis of rapidity \cite{KPH01} and 
transverse momentum \cite{HZ02} dependencies of medium effects on the
final spectrum. Regarding (anti-)deuterons, collective flow patterns emerge
naturally as they are built-in both in formation and dissociation rates,
providing a useful probe of the freeze-out properties of the produced
medium \cite{M99}.
Details of how different effects play a role both for quarkonia and
for (anti-)deuterons can only be addressed with sufficiently realistic
numerical studies. In fact, a detailed knowledge of the medium
constituents' and of the transition probabilities is required. In the
following we will limit our discussion to a simple example to
clarify the physical content of the result contained in eq.~(\ref{sollocal}).

Before doing so we recall that the invariant spectrum of bound states can be
computed by means of the Cooper-Frye formula \cite{CF74}.
Given a hyper-surface $\tau_f = \tau_f(\eta,\vec r_\perp)$, one has
\be
\frac{dN_B}{dy\,d^2\vec p_\perp} = 
\int_{\tau_f}\!\frac{d\sigma_\mu p^\mu}{(2 \pi)^3}
\ \fb\left(y,\vec p_\perp,\eta,\vec r_\perp,\tau_f(\eta,\vec r_\perp) 
\rule{0pt}{10pt} \right)
\ee
where $\fb$ is given by eq.~(\ref{sollocal}) and
\bea
d\sigma_\mu p^\mu \!&=\!& d^2\vec r_\perp d\eta
\, \left[ \tau_f\, m_\perp \cosh(y-\eta) \rule{0pt}{12pt} \right.\\
& & \left. \hspace{1cm} - \ \tau_f\, \vec p_\perp \cdot 
\frac{\partial \tau_f}{\partial \vec r_\perp} - m_\perp 
\frac{\partial \tau_f}{\partial \eta}\sinh(y-\eta)\right] \nonumber
\eea
is the invariant volume integration measure.

To discuss the physical content of the solution given with 
eq.~(\ref{sollocal}), or equivalently with eq.~(\ref{solglobal}), we 
focus on a single aspect of it. We refrain from discussing the part
describing dissociation of initially present bound states. This was
extensively done in \cite{BO89} for charmonium and only specific models
of the medium can give qualitatively new answers. On the other hand
the part describing formation deserves more attention. As mentioned 
earlier it can potentially account for quarkonium formation in a QGP,
as suggested in \cite{TSR01} and become the dominant mechanism of
production at high enough energies (RHIC, LHC). Moreover, the second
term should describe formation of (anti-)deuterons in the late stages
of the collision. While much work has been done in this domain, efforts
were limited to applications of the coalescence model at freeze-out
(See \cite{PBM98,SH99} for some recent results). Only recently \cite{ISZ03}
a new attempt was made to address this problem. 

Therefore, focussing attention on bound state formation in the medium, we 
neglect the first term in eq.~(\ref{sollocal}) either because dissociation
destroys all the initially present bound states or because there are no
bound states to start with. With significant simplifications of the solution 
in order to follow the essential physics, we neglect all dependencies on
$y,\vec p_\perp,\eta$ and $\vec r_\perp$. The remaining solution can be
written as
\be
N_B(\tau) = \int_{\tau_0}^\tau\!\!d\tau' \, \Lambda_F(\tau')
\, \exp\left[- \int_{\tau'}^\tau\!\!d\tau'' \, \Lambda_D(\tau'')\right]\,.
\label{solsimple}
\ee
This situation is equivalent to having solved the first order ordinary 
differential equation $dN_B/d\tau = \Lambda_F - \Lambda_D\,N_B$ with 
initial condition $N_B^0 = 0$.
We then assume that the time dependence of the formation and dissociation
rates are
\be
\Lambda_F(\tau) = \frac{P_F\,N_{c_1}N_{c_2}}{\tau}\ \ \ \ \ {\rm and}
\ \ \ \ \ \Lambda_D(\tau) = \frac{P_D\,N_D}{\tau}\,,
\ee
where $P_{F,D}$ are formation and dissociation probabilities, $N_{c_{1,2}}$
indicates the number of particles which can form the bound state (either
$pn$ or $q \bar q$ pairs) and $N_D$ is the number of particles that can
dissociate the bound state. 
With the latter assumption one can readily perform the time integrations
in eq.~(\ref{solsimple}) obtaining the final number of bound states at 
$\tau_f$ as
\be
N^f_B = \frac{P_F\,N_{c_1}N_{c_2}}{P_D\,N_D}
\,\left[ 1 - \left(\tau_0/\tau_f\right)^{P_D N_D}\right]
\propto \frac{N_{c_1}N_{c_2}}{N_D}
\label{solsimple2}
\ee
when $P_D N_D \gg 1$. This last results indicates that the final yield
is proportional to the number of particles which can form the bound state
and inversely proportional to the number of particles that can
dissociate it. With the help of numerical simulations, the consequences
of this scaling law for deuterons were discussed previously in \cite{SNK95}.
Moreover, since the number of dissociating particles is proportional
to the volume of the medium, either at hadronization for quarkonia or at
freeze-out for (anti-)deuterons, we recover the well known coalescence
formula \cite{CK86}, here newly obtained with dynamical considerations. 
The above simplified derivation gives some insight in 
the coupled dynamics of dissociation and formation. In particular, as
discussed in \cite{GKMSG02,T03,GG99}, the experimental observation of an 
inverse
proportionality of the number of quarkonia with the total number of
produced hadrons, might give support to this new mechanism of production
in a QGP. 

We can now conclude by summarizing the results presented above. The aim of
this paper has been twofold: to present a unified theoretical description 
of bound state formation in the strongly interacting medium, formed with a
heavy-ion collision at high energy, and to provide a useful tool to be
applied in numerical calculations of bound state spectra. This was achieved
by means of a kinetic equation with a collision term consisting of formation
and dissociation parts. The equation was solved exactly in closed form and the
result given in eqs.~(\ref{solglobal}) and (\ref{sollocal}). 

The physical content of the solution was discussed and a simplified version
of it was given with eq.~(\ref{solsimple2}) in order to illustrate that the
yield of bound states form in the medium satisfy the scaling embedded in the
coalsecence formula, {\it i.e.} direct proportionality with the number of
bound state constituents and inverse proportionality with the system volume.
The full solution, on the other hand, provides a convenient tool for future
numerical studies, which require quantitative knowledge of the phase-space
distributions of the degrees of freedom involved and of the transition
probabilities among those degrees of freedom.

\section*{Acknowledgements}
I am grateful to R.~Schneider and W.~Weise for illuminating discussions
and to T.~Renk for a critical reading of the manuscript. 
This work was supported in part by BMBF and GSI.

\end{document}